\documentclass{eas}
\usepackage{graphicx}
\newcommand{\HII}{H\,{\sc ii}}

\newcommand{\Bg}{Br$\gamma$}

\newcommand{\Myr}{$M_{\odot}$~yr$^{-1}$}

\newcommand{\kms}{km~s$^{-1}$}

\newcommand{\vinfty}{$v_{\infty}$}
\newcommand{\Md}{$\dot{M}$}

\newcommand{\Teff}{$T_{\rm{eff}}$}
\begin{document}

\title{Massive Stars in Transition} 
\author{Paul A. Crowther}\address{Dept. of Physics \& Astronomy\\
University College London\\Gower Street\\London WC1E 6BT}
\runningtitle{Crowther: Massive Stars in Transition}
\begin{abstract}
We discuss the various post-main sequence phases of massive stars,
focusing on Wolf-Rayet stars, Luminous Blue Variables, plus connections
with other early-type and late-type supergiants. 
End states for massive stars are 
also investigated, emphasising connections between Supernovae originating from
core-collapse massive stars and Gamma Ray Bursts.
\end{abstract}
\maketitle
\section{Introduction}


Material introduced in my article dealing with OB stars is built upon 
here for post-main sequence phases of massive stars,
namely Wolf-Rayet stars, slash stars, Luminous Blue Variables (LBVs), B[e]
supergiants and cool supergiants. What are the properties and interrelations
between these extreme objects? What happens to these stars at the end of
their short lives?

The most dramatic explosions in the Universe, Gamma Ray Bursts (GRBs), 
convert a sizable fraction of a Solar Mass into energy within seconds.
The currently favoured scenario for most GRBs is via the collapse of a
massive star to form a blank hole, with a corresponding Supernova explosion.
Closer to home, $\eta$ Car, the archetypal LBV is
one of the most extreme Milky Way stars and briefly was the second
brightest star in the night sky in the middle of the 
19th century, despite  lying 8000 light years from the Sun. 
During this eruption it ejected material at an {\it average} 
rate of one Earth mass every 15 minutes for two decades.
Wolf-Rayet (W-R) stars, the evolved descendants of massive OB stars  
throw off several solar masses over lifetimes of 10$^5$ yr.
W-R stars, plus their subsequent SN explosions are major
contributors to 
the chemical enrichment of galaxies (Leitherer et al. 1992; Dray et al. 2003).
It is apparent
that the advanced stages of the most massive stars represent exceptional
astrophysical objects, which are discussed below.


\section{WR photometric systems}

The standard Johnson $UBV$ filter set provides adequate
continuum measurements for Luminous Blue
Variables and slash stars. However, the intrinsic colors of W-R stars are 
complicated by their strong emission lines. 
These features greatly modify the flux
measured in each optical filter, especially those of the relatively broad
band $UBV$ system. Such measurements may overestimate the true
continuum level by up to 1 magnitude in extreme cases, or more typically
0.5 mag for single early-type W-R stars. Consequently,
Smith (1968b) introduced {\it narrow band} $ubv$
filters specifically designed for W-R stars to minimize the effect of
emission lines (but their effect cannot be entirely eliminated). Most
photometry of W-R stars has used this  $ubv$ filter system. 

The following
relations relate the broad and narrow band optical indices for W-R
stars (from van der Hucht 2001).
\begin{eqnarray*}E_{B-V} & = & 1.21 \, E_{b-v}\\
A_v & = & 4.1 \, E_{b-v} = 1.11 \, A_V \\
(B-V)_0 & = & 1.28 \, (b-v)_0 = 0.11 \\
M_V & = & M_v + 0.1 
\end{eqnarray*}

A recent compilation of the spectral subtype dependence of $(b-v)_0$
for W-R stars (van der Hucht 2001) reveals values in the range $-$0.2 to $-$0.35
for WN and WC stars, with the possible exception of WC9 stars.
Observations in the $ubv$ system are mostly used to
determine the interstellar extinction, $A_v$, rather than to be able to
say anything about the star itself.

As with OB stars, absolute magnitudes of slash stars and 
W-R stars rely on calibrations
of stars that are members of Galactic or Magellanic Cloud
clusters or OB assocations. Confirmed numbers of LBVs are insufficient
for absolute magnitudes to be useful. Only
$\gamma$ Velorum (WC8+O), the only naked eye W-R star
visible (from the southern sky) was close enough for a reliable 
Hipparcos measurement. 

 van der Hucht (2001) also provides a modern $M_{v}$-spectral type
calibration for WN and WC stars. Absolute visual magnitudes for single
W-R stars range from $M_{v}=-2.5$ for early WN and WO stars, to
$M_{v}=-4.5$ for late-type WC stars, and
$M_{v}=-7$ for late-type WN stars. There are no reliable 
bolometric corrections published for W-R stars to date, 
although individual values range
from $-$2.7 amongst very late WN stars (Crowther \& Smith 1997), 
to approximately $-$6 for weak-lined, early-type WN stars (Crowther et al. 1995c)
and WO stars (Crowther et al. 2000). For LBVs and slash stars, bolometric
corrections closely mimic those of corresponding OB supergiants (Crowther 1997).



\section{Spectral Classification}

\subsection{W-R stars}

Spectral classification of W-R stars is derived from emission line 
strengths and line ratios (Beals \& Plaskett 1935). 
The modern system for both nitrogen-rich (WN) and carbon-rich (WC)
stars is based upon that introduced by Smith (1968a). 
Details of the classification scheme are  contained in 
van der Hucht (2001) where an up-to-date catalogue of Galactic 
W-R stars is also given. 

WN spectral subtypes utilize a one-dimensional scheme, involving
line ratios of N\,{\sc iii-v} and He\,{\sc i-ii} lines, 
ranging from WN2 to WN9, extended to later 
subclasses by Smith et al. (1994) to accommodate cool slash stars
(see below). At the earliest classical subtypes 
solely He\,{\sc ii} and N\,{\sc v} are present, with He\,{\sc i} and lower ionization
stages of nitrogen absent, whilst at the latest subtypes, He\,{\sc i} is strong,
He\,{\sc ii} present but weak, N\,{\sc iii} present, with higher ionization
stages absent. Since the definition of some subtypes  (e.g. WN8) involves
the relative strength of nitrogen to helium emission lines, there may be
an inherent metallicity dependence. Consequently, Smith et al. (1996) 
devised a scheme that involved ratios of lines from either helium or
nitrogen.  
Various multi-dimensional systems have been proposed, generally involving
line strength or width, none of which have been generally adopted, 
the most recent of which was by Smith et al. (1996), including
criteria for the presence of hydrogen (h), and broad (b)
line widths.

WC spectral subtypes depend on the line ratios 
of C\,{\sc iii} and C\,{\sc iv} lines along with the appearance of O\,{\sc v} (Smith
et al. 1990), spanning WC4 to WC9 amongst massive stars 
\footnote{Some H-deficient central stars of Planetary Nebulae show
carbon-sequence W-R spectral features, that 
extend to later spectral types (Crowther et al. 1998).}, which has recently
been revised to account solely for C\,{\sc iii-iv} lines (Crowther et al. 1998).
Oxygen-rich WO stars form an extension of the WC sequence to higher
excitation, exhibiting strong O\,{\sc vi} emission (Barlow \& Hummer 1982). The
most recent scheme involves WO1 to WO4  depending on the relative
strength of O\,{\sc v-vi} and C\,{\sc iv} emission lines (Crowther et al. 1998). 
Although line widths of early WC and WO stars are 
higher than late WC stars, width alone is not a defining criterion
for each spectral type. 

As with O stars, atlases of W-R spectra are available at
ultraviolet (Willis et al. 1986) and near-IR (e.g. Figer et al. 1997) wavelengths, 
permitting approximate spectral types to be assigned. 


\subsection{Slash Stars}    \label{subsection_trans}

As described in my first article, 
a subset of O stars exhibit emission lines
in their optical spectra. Although these are generally much weaker
than in W-R stars, 
there are a number of stars with intermediate characteristics, 
which have been coined `transition' objects between O-type and W-R.
There are two flavours, hot and cool `slash' stars. 

\subsubsection{O3If/WN Stars}

Walborn (1973) discussed the close morphological
relationship between various early-type stars in Carina, namely
HD~93129A (O3~If, recently revised to O2~If by Walborn et al. 2002) 
and HD\,93162 (WN6ha).
Subsequently, stars with intermediate characteristics were identified
in the LMC, such as Sk$-67^{\circ}$~22 and Melnick 42, and
assigned spectral types of O3\,If/WN (Walborn 1986).
Whatever the nomenclature issues, the important point is that
merely a small change in wind density
distinguishes the most extreme O2--3 supergiants 
from the least extreme WN stars, such as those identified
in  Carina, R136 and NGC\,3603. These WN stars 
are hydrogen-rich (h) and contains (apparently) intrinsic
absorption (a) lines (Crowther et al. 1995b; Crowther \& Dessart 1998),
such that they have apparently more in common with early O supergiants
than
the bulk  of hydrogen-free, core He-burning W-R stars.

For completeness,
cooler analogues of the hydrogen-rich Carina WN stars have also been identified.
A careful reconsideration of the spectral types of three Galactic O6--8\,Iafpe stars led 
Bohannan \& Crowther (1999) to revise their classifications to WN9ha.
They provided provide suitable criteria for consideration of previously undiscovered
stars with similar spectral morphologies.

\subsubsection{Ofpe/WN9 Stars} 

Walborn (1977, 1982) identified a group of late O
LMC emission line stars as Ofpe/WN9 subtypes,
with apparently intermediate characteristics of Of and late WN stars. 
Members of this group were
 re-classified WN9, 10 or 11 by Smith et al. (1994) and Crowther \& Smith 
(1997) on the basis that characteristic O star
photospheric lines were absent. Additional examples have been identified in
M33, IC~10 and NGC\,300.   WN11 subtypes closely resemble 
extreme early-type B supergiants,  except for the presence of 
He~II $\lambda$4686 emission. Interest in this group of very late WN stars
grew because of their high frequency of associated H~II regions, plus 
an apparent connection with LBVs. One prototypical
Ofpe/WN9 star (subsequently re-classified as WN11), R127 in the LMC, was
later identified as a LBV. Conversely, AG~Car, a well known Galactic LBV,
exhibited a WN11-type spectrum at extreme visual minimum (Smith et al. 1994).

In the $K$-band, WN9--11 stars exhibit strong He\,{\sc i} 2.058$\mu$m and \Bg emission lines
(Figer et al. 1997; Bohannan \& Crowther 1999). Observations obtained within the
last decade reveal that a large number of stars 
within clusters at, or close to, the Galactic Centre also show this spectral morphology.
As in the optical, very late WN stars also resemble early B supergiants, such as
 P Cygni. Consequently, the presence of He\,{\sc ii} near-IR lines is required to
confirm these stars as bona-fide WN stars, which is currently lacking in most cases.
Examples are shown by  Najarro et al. (1997a) for He\,{\sc i} sources in 
the central cluster, Figer et al. (1999a)  in the Quintuplet cluster
Emission-line stars located in the Arches cluster most closely resemble hot
slash stars or weak-lined WN stars (Blum et al. 2001).

\subsection{Luminous Blue Variables}  \label{LBVs}

Observationally,
LBVs inhabit a part of the H-R diagram that is bereft of normal stars. This
region is known
as the Humphreys-Davison limit (Humphreys \& Davidson 1979), above
which either no supergiants are observed, or their transit through this
region is so brief that none would be expected on statistical grounds. 
The photometric and spectroscopic 
variability of LBVs is suspected of being so large since they stars lie 
close to the Eddington limit. Beyond this limit, stars would not remain stable.

In contrast with O and W-R stars, there is no single spectroscopic
defining characteristic to define a LBV; historically it simply 
had to be a variable, luminous, blue star (Conti 1984). Examples are
known in our Galaxy, including $\eta$ Carinae and P Cygni, the Magellanic Clouds, 
known collectively as S Doradus-type variables, plus M31 and M33, these
known historically as the Hubble-Sandage variables, after their discoverers
in the 1950's. More recently, it has
been proposed that the presence of an associated \HII region, a 
remnant from recent mass-losing episodes, may be a defining characteristic. 

LBVs are generally restricted to 
those stars with substantial photometric variability on timescales of
decades, plus associated spectroscopic changes. 
Two types  of photometric variability are commonly discussed.

\subsubsection{LBVs during quiescence}

LBVs exhibit micro-variations ($\sim$0.1 mag) over short timescales, and
irregular variations of up to 1--2 mag, in which the spectral type may
vary between an early B supergiant at visual {\it minimum} 
to a late B or early A supergiant at visual {\it maximum}. Indeed, 
some LBVs have even proceeded to spectral types as late as 
F or even G.   Classic examples of irregular variability 
over the past few decades 
include  AG~Car and HR~Car in the Milky Way, R~71 and R~127 in the LMC.

Because of the irregular
nature of variability, it is almost certain that many `dormant' LBVs
await discovery, with all luminous blue supergiants potentially undergoing
a LBV outburst at some point in their evolution. 
A fine example is HDE\,316285, an early
B supergiant with a more extreme wind than P~Cyg and spectroscopic
similarities to $\eta$ Car (Hillier et al. 1998). Only 
moderate spectroscopic variability has been observed to date, 
such  that  an LBV nomenclature is  not (yet) appropriate.


\subsubsection{Giant eruptions}

More remarkably, giant eruptions in LBVs have been observed, 
in which photometric variability exceeds 2 mag in the visual. 
Such outburts are exceedingly rare, with only P Cygni and $\eta$ Carinae 
known to have undergone
such an eruption between the 15th and late 20th centuries. 
It is almost certain that other stars will have undergone such eruptions
during this time, but were not sufficiently bright to have been noticed 
by astronomers. P Cygni was discovered in 1600 by the Dutchman
Bleau, when it suddenly appeared as a naked eye star 
(no star was apparent before). It remained bright
for many years, before fading and re-appearing in 1655, and then 
fading again. At present it is around 5th magnitude, and so is at present the
brightest LBV in the sky. P~Cygni profiles, generally observed in 
UV resonance lines in hot star winds are described as such because
P~Cygni displays these profiles in hydrogen and helium lines in
its optical spectrum. 

During the early 1990's, two other stars have erupted as LBVs -- 
the well known WN+O binary (or perhaps triple) system
HD~5980 in the Small Magellanic Cloud  (Barba et al. 1995), plus a previously 
unknown star in the GH~II region NGC~2363 of the 
low-metallicity galaxy NGC~2366, labelled V1 (Drissen et al. 2001). 

\subsection{B[e] supergiants}

The term `B[e] star' was coined by Conti (1976), and designates those B
stars which show forbidden (`[]') emission (`e') lines in their optical spectrum.
B[e] stars should not be confused with main-sequence Be stars.
B[e] supergiants, or sgB[e] following Lamers et al. (1998),
are extremely luminous massive stars observed in 
the Magellanic Clouds that are distinguished by a plethora of 
low-excitation emission lines including (broad) Balmer lines 
and (narrow) singly ionized metals, including permitted and forbidden iron lines.
These stars also show a significant near- and mid-IR dust excess, providing evidence
for hot, circumstellar dust, such that
the standard interpretation for these unusual properties
is an equatorial excretion disk, plus a fast polar stellar wind (Zickgraf et al. 1985).
In common with other luminous supergiants, B[e] supergiants appear to be nitrogen
rich, showing evidence for the products of the CN-cycle at their surfaces.
Gummersbach et al. (1995) provide a thorough discussion of the luminosities and
effective temperatures of the known B[e] stars in the Magellanic Clouds.


B[e] supergiants occupy a similar location in the H-R diagram to the LBVs, although
a firm connection is currently lacking (Conti 1997). It is likely that
such stars are not post-red supergiants, since angular momentum would be lost
during this phase which would not be regained once the stars return blueward.
Photometric variability of B[e] supergiants is generally small, although
the binary system R4 in the SMC has shown an LBV-type photometric and
spectroscopic outburst of 0.5 mag in
the mid-1980s (Zickgraf et al. 1996). Hen S18, another SMC B[e] supergiant
is also spectroscopically variable (Sanduleak 1977)\nocite{SAN77}, 

\section{Circumstellar nebulae}

The recent mass-loss history of a subset of hot, luminous stars may be
studied their circumstellar environment. 


$\eta$ Car, at the heart of the great Carina nebula, is known to have 
shed approximately 3M$_{\odot}$ during a couple of decades between 1837--1860.
During that time it became one of the brightest stars in the southern sky
and so has one of the most remarkable photometric histories of any naked-eye
star. The event remains unprecedented to this day, and corresponds 
to losing the equivalent mass of the Earth every 15 minutes for 2 decades.
Such a large amount of material should be visible. Indeed it is, and has
been named the Homunculus nebula.  HST/WFPC2 images of the Homunculus reveal
incredible bi-polar emission lobes (Morse et al. 1998). 
Its orientation on the sky is such that
the central star(s) is largely masked from view. 
Damineli et al. (1997, 2000) found evidence in favour of a 5.5 year period
for $\eta$ Car involving two massive $\sim$70$M_{\odot}$ stars.
 
Nevertheless, we do see evidence that  other LBVs underwent similar eruptions
through their circumstellar environment. The ejecta nebula 
associated with the famous Luminous Blue Variable AG Car 
probably differs only from the Homunculus Nebula through a much greater 
dynamical age and orientation (Langer et al. 1999) . The dynamical age of a nebula, 
$\tau_{\rm dyn}$,  is 
simply the radius of the nebula, divided by the
expansion velocity. The size of the nebula associated with AG~Car 
is 0.6~pc (assuming a distance of 6~kpc), whilst the expansion velocity
is 70~km\,s$^{-1}$ such that $\tau_{\rm dyn}$=8,400 yr, and is typical
of other LBVs, except for $\eta$ Car, for which its dynamical age is in
good agreement with the 1837 event.

The Pistol Star at the Galactic Centre represents the best example 
of an early-type star that has not been observed to dramatically vary
in continuum flux, but possesses a \HII region (Figer et al. 1998, 1999b),
a characteristic which has been used to classify it as an LBV.

Ejecta \HII regions are also associated with a subset of Wolf-Rayet stars,
predominantly amongst WN stars. Beautiful examples include
 M1~67 (Grosdidier et al. 1998) and NGC\,6888 (Moore et al. 2000).
It is likely that these were ejected during a
previous (LBV?) phase, but are currently photo-ionized by the W-R star.
Other W-R stars possess associated \HII regions, albeit composed of
material exhibiting ISM abundances (Esteban et al. 1992).
 

\section{Stellar Properties}

As discussed in my first article relating to OB stars, spectral line 
techniques, as opposed to continuum techniques, 
are required to derive stellar properties of W-R and other early-type
mass-losing stars. Such rules apply equally to LBVs during their
A and B-type spectral phases, since they
too possess strong stellar winds, and their surfaces are probably
H-depleted. 

Spherically extended, non-LTE line blanketed 
model atmospheres are essential for the detailed analysis of
Wolf-Rayet stars, slash stars and Luminous Blue Variables, of which
{\sc cmfgen} (Hillier \& Miller 1998) and that developed 
by the Potsdam group (Gr\"{a}fener et al. 2002) 
have been most widely used to date. 

\subsection{Temperatures}

As with normal O stars, spectral lines of helium (or nitrogen)
may be used to derive the  \Teff\ of Ofpe/WN9 stars
and WN stars, whilst silicon diagnostics offer the possibility of
determining temperatures of LBVs during their B supergiant phases.
Carbon or oxygen lines can be used to diagnose WC and WO stars.
Stellar temperatures for spherically extended 
stars are more difficult to characterize than dwarfs,
since the geometric extension in these stars is comparable with the stellar
radius particularly for W-R stars and LBVs. 
In practice, one defines a radius as the distance from the center
where the mean Rosseland optical depth,
$\tau_{\rm Ross}$, is of order 10. For normal stars
this does not differ from the usual radius at $\tau_{\rm Ross} = 2/3$,
but differs greatly for W-R stars, and represents a much more representative
ionization indicator. In extreme cases, such as $\eta$ Car, the high density
of the wind completely veils the stellar photosphere such that it is 
impossible to characterize a stellar temperature (Hillier et al. 2001).

We have already discussed that the recent incorporation of line blanketing
in analysing the {\it photospheric}
 spectrum of OB stars has led to a substantial
{\it decrease} in derived temperatures (Martins et al. 2002; Crowther
et al. 2002b). 
In contrast, similar advances for W-R stars has led to an {\it increase}
in derived temperatures obtained from {\it emission} lines  (Herald et
al. 2001).  Stellar temperatures of W-R stars
range from 30,000K amongst late-type WN stars to well in excess of 
100,000K for early-type WN stars and WO stars, 
although differences in wind density complicate accurate determinations.
Hot slash stars and H-rich WN stars have stellar temperatures at the
high end of the O main-sequence, typically $\sim$40--45kK. These stars
have the highest luminosities of all `normal' stars.

Increased stellar temperatures of W-R stars relative to OB stars
implies that their contribution to the Lyman continuum of young stellar 
clusters in increased. 
However, since blanketing redistributes extreme-UV
radiation to lower energies, predicted fluxes at higher energies, such
as those shortward of the He\,{\sc ii} 
$\lambda$228 edge, are actually reduced (Smith et al. 2002).
It is solely hot W-R stars with relatively weak winds that are a significant
source of He\,{\sc ii} Lyman continuum radiation.

Luminous Blue Variables range in stellar temperatures between $\sim$8kK 
at visual maximum, when they typically exhibit an A-type hypergiant spectrum, 
and 25kK at visual minimum, with an early B subtype. 
Consequently these stars
are minor contributors to the Lyman continuum photons of young clusters.
Because of the change in B.C. with \Teff, 
visual changes in LBVs are generally considered to occur at fairly {\it constant} 
bolometric luminosity (Crowther 1997). 
It is only for LBVs undergoing giant eruptions,
such as $\eta$ Car in the 19th Century, that this is not true.
Finally, WN9-11 stars, possess temperatures in the range 25--30kK,
filling-in the temperature sequence between LBVs and W-R stars
(Smith et al. 1994; Crowther \& Smith 1997).

\subsection{Masses}
 
Spectroscopic measurement of masses in post-main sequence, early-type
stars is possible in principle via  analysis of line wings. However,
stellar winds in most cases affect   line spectra sufficiently to hinder 
gravity determinations. Similarly, evolutionary masses are often 
complicated by the uncertain (blueward or redward) evolutionary
state of the star. Estimates of initial masses can be obtained from
cluster turn-offs (e.g. Massey et al. 2001; St-Louis et al. 1998).

In the case of Wolf-Rayet stars and Luminous Blue Variables, 
one relies on direct measurements of current masses via analysis of binary orbits.
Amongst LBVs, only R81 in the LMC is a binary with a well
characterized orbit (Tubbesing et al. 2002), whilst several examples are known 
amongst W-R stars, widely studied by V. Niemela and colleagues.
By way of example, the binary nature of $\gamma$ Velorum permits a  
mass determination for both the W-R and O components, with
 10$M_{\odot}$ obtained for the WC component. Fortunately, 
there is predicted to 
be a robust relationship between the mass of a hydrogen-free WN or WC
star and its stellar luminosity (Schaerer \& Maeder 1992).
For $\gamma$ Vel, good agreement is found between the direct 
measurement  and that resulting from a recent line-blanketed
spectroscopic analysis (De Marco et al 2000).

Not all W-R stars have masses which are relatively low. 
The eclipsing binary
Carina WN7ha+O system HD\,92740 has provided a exceptionally high
mass for the WN component of $\sim$55$M_{\odot}$ (Schweickhardt
et al. 1999). This result, together
with spectral analyses revealing that this star is rich in hydrogen 
has led to a subset of W-R stars being dubbed H-burning
main-sequence O stars with particularly strong stellar winds.
Unfortunately, direct information on the processes going on within the
core of a massive star are very hard to determine without recourse to
evolutionary models.

\section{Stellar Wind Properties}

As for OB stars, several techniques may be used to study the wind
properties of W-R stars and LBVs. These stars have the advantage
that their more powerful winds permit velocity and mass-loss rate
determinations other than in the UV and at H$\alpha$.
Analytical relationships (Wright \& Barlow 1975) permit measurements
of mass-loss rates of LBVs and W-R stars via IR-radio continua, providing
distances and wind velocities are known. Current sensitivities of 
radio telescopes restricts this method to stars within a few kpc of 
the Sun. 

\subsection{Wolf-Rayet stars}

In addition to the usual ultraviolet P~Cygni wind velocity
diagnostics (Prinja et al. 1990), 
optical, and especially near-IR P~Cygni lines of He\,{\sc i} 
permit accurate
wind velocities for W-R stars (Howarth \& Schmutz 1992; Eenens
\& Williams 1994). ISO mid-IR fine structure lines also facilitate 
direct measurement of wind
velocities in some cases. Wind velocities of WR stars increase
from late, to early- spectral types. WN8 and WC9 stars possess
wind velocities of order $\sim$1000 \kms, whilst WN3 and WC4 stars
have winds on average a factor of three times larger. WO stars
possess the highest wind velocities of any (stable) star, up to
5500 \kms (Kingsburgh et al. 1995).

Radio mass-loss rates of W-R stars have been derived by
Leitherer et al. (1997) using the Wright \& Barlow (1975) analytical relationship,
although care needs to be taken with regard to the elemental 
abundances and ionization balance in the radio emitting region.
This method is distance limited, due to the limited sensitivity of current
radio telescopes, and one should obtain radio fluxes at more than one wavelength
in order to avoid overestimates of mass-loss rates due to non-thermal emission
in a close binary (Chapman et al. 1999). Non-LTE model analysis of 
UV, optical and IR  line profiles is more widely used for Galactic and extragalactic W-R stars
(e.g. Dessart et al. 2000). 

Unblended spectral regions are scarce in the ultraviolet and visual, particularly
for WC stars, such that synthesis of  fluxed spectroscopy is the 
preferred method. It quickly became apparent that it was
impossible to reproduce both peak line intensities -- the strength of which
are proportional to the square of the density -- and red electron
scattering winds -- linearly proportional to the density -- under
the assumption of homogeneity (Hillier 1991). Many other observations
(Lepine et al. 2000) 
 and theoretical arguments (Dessart \& Owocki 2002) 
point to severely clumped winds in W-R stars, although precise volume
clumping factors are challenging to derive. 

Most observations are
consistent with clumping factors of $f\approx$0.1, in the sense
that clumped mass-loss rates are a factor of $1/\sqrt{f}\approx$3
times lower than previous homogeneous rates (Hamann \& Koesterke 1998). 
Clumped W-R
mass-loss rates typically span a range from 1--3 $\times$ 10$^{-5}$ \Myr,
such that wind performance numbers,  {\Md \vinfty\,/(L/c)
the factor by which the wind exceeds the single scattering limit, 
lie in the range 1--10. Nugis \& Lamers (2000) provide
calibrations for WR mass-loss rates as a function of luminosity,
He and metal content.

\begin{figure}
\includegraphics[height=7.3cm]{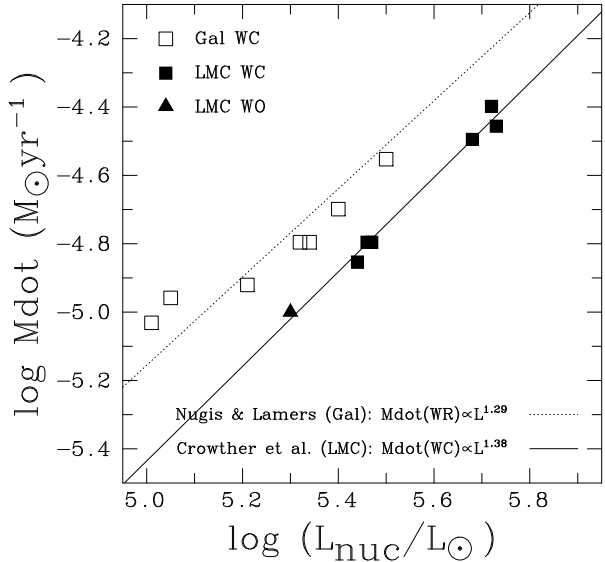}
\caption{Comparison between (clumped) mass-loss rates
of Galactic (open symbols) and LMC (filled symbols) WC
stars, revealing systematic weaker winds in the latter,
by an amount predicted by radiatively driven winds. The
solid line is a fit to the LMC data (Crowther et al. 2002a),
whilst the dotted line is a calibration from Nugis \& Lamers
(2000).}
\end{figure}

\subsection{Metallicity dependence for WR winds?}

In contrast with normal OB stars, it is not yet established that 
W-R winds are (solely) driven by radiation pressure. A metallicity
dependence of O star winds is well established theoretically (e.g.
Vink et al. 2001), for which observational support has been demonstrated
(Puls et al. 1996). 

In the case of W-R stars, 
derived mass-loss rates of Milky Way WN stars span a 
wide range, such that comparisons with Magellanic Cloud WN stars 
can be interpreted in a variety of ways (Crowther \& Smith 1997; 
Crowther 2000). In contrast, the wind densities of WC stars form a
much cleaner distribution. Crowther et al. (2002a) demonstrated that
singe early-type 
WC stars in the LMC possessed winds that were weaker
than those of comparable Milky Way stars. This difference, 
identical to that expected from radiatively driven wind theory for 
OB stars, led
Crowther et al. (2002a) to conclude that WC spectral types result
primarily from metallicity differences, rather than abundances as
suggested by Smith \& Maeder (1991). 


Rotation is much more 
difficult to measure in W-R stars than in OB stars, since photospheric
features are lacking. Attempts have been made, but without much success.
Spectropolarimetry does permit information to 
be obtained regarding geometry, which indicates that most Galactic W-R
stars are spherical, with a few notable exceptions amongst WN stars 
(Harries et al. 1998). 

\subsection{Slash stars}

Hot slash stars (O3If/WN stars) have 
wind properties intermediate between O2--3 supergiants, 
and weak-lined, hydrogen rich WN5--7 stars, i.e. high
wind velocities of $\sim$3000 \kms, plus mass-loss rates
approaching 10$^{-5}$ \Myr (de Koter et al. 1994).
In contrast, cool slash stars (WN9--11 stars) possess relatively
low wind velocities, generally $<$500 \kms, plus moderate
mass-loss rates (Crowther et al. 1995a; Crowther \& Smith 1997). 
Galactic Centre slash stars 
have higher wind velocities, in the range 500--1000 \kms
(Najarro et al. 1997a).

\subsection{Luminous Blue Variables}

Typically, LBVs during quiescence
possess wind velocities which are a factor of several
times lower than corresponding OB supergiants, 
with mass-loss rates a factor of several times higher  (Crowther 1997).
For example, P Cygni has a wind velocity of 185 \kms, in contrast to 
a typical $v_{\infty}$=1060 \kms for normal B1 supergiants 
(Prinja et al 1990).
Leitherer (1997) studied the mass-loss variability of AG Car during
its excursions around the H-R diagram, deducing a `random walk' in
which the mass-loss rate did not undergo dramatic variations. R71
appears to be unusual in possessing a much reduced mass-loss rate
at visual minimum (Leitherer et al. 1989).

In contrast,  LBVs undergoing a major eruption possess exceptional
wind properties. $\eta$ Car is believed to have shed in excess of 
3$M_{\odot}$ over  between 1840--1860, such that its average mass-loss
rate during that event was $\sim$10$^{-1} M_{\odot}$ yr$^{-1}$. Even
today, during its quiescent phase, its mass loss rate is perhaps only
a hundred times lower. Note that a giant eruption does not necessarily 
imply exceptional mass-loss.  NGC\,2363-V1 is an LBV that is currently
undergoing a giant eruption with a mass-loss rate that is comparable
to $\eta$ Car during {\it quiescence} rather than {\it eruption} 
(Drissen et al. 2001).

\section{Abundances}

\subsection{Wolf-Rayet stars}

It has long been suggested that W-R abundances represented the
products of core nucleosynthesis.  However, it was only with 
Balmer-Pickering decrement studies (Conti et al. 1983) that
 indeed hydrogen was found to be severely depleted in all WR stars. 
Nitrogen and carbon abundances are consistent with CNO cycle values 
in WN stars from both atmospheric and nebular analyses, with nitrogen
representing up to $\sim$1\% of the surface mass. In WC and WO stars 
extremely high carbon and oxygen abundances are found, typically in excess of 
50\% by mass, clearly reflecting both triple--$\alpha$ and $\alpha$--capture products. 


Recombination
line studies (Smith \& Hummer 1988; Smith \& Maeder 1991), 
using theoretical coefficients for different transitions 
are most readily applicable to WC stars, since they show a large number of 
spectral lines in their optical spectra. Atomic data is most readily available 
for hydrogenic ions, such as C\,{\sc iv} and O\,{\sc vi}, so early-type WC and WO stars
 can be most easily studied. In reality, optical depth effects play a role too. 
Limitations with earlier approaches have greatly been overcome with
modern spectroscopic non-LTE studies. WN 
abundance patterns are consistent with material processed by the CNO nuclear
burning cycle, whilst WC stars reveal products of He burning. 

A clear subtype effect in the H--abundance is found for WN stars, with
late-type WN stars generally showing some hydrogen, with early-type WN stars
hydrogen free. Exceptions to this general rule do exist, such as HD\,177230
(WN8o, Crowther et al. 1995b) and HD\,187282 (WN4(h), Crowther et al. 1995c). 
WNLha stars are rich in hydrogen (Crowther \& Dessart 1998).
Non-LTE model atmosphere analyses indicate fully CNO processed material
at the surfaces of most WN stars (e.g. Crowther et al. 1995b).

Conti \& Massey (1989\nocite{CM89})  identified 7 Galactic and 
2 LMC stars as having intermediate WN--WC characteristics based on 
anomalously high C\,{\sc iv}$\lambda$5801/He\,{\sc ii}$\lambda$4686 line 
strengths compared to normal WN stars. Modern analyses of these `transition'
WN/C stars indicate elemental abundances intermediate between WN and WC stars,
with N/C$\sim$1 by number (Crowther et al. 1995d), supporting rotationally
induced mixing in W-R stars.

Neither hydrogen nor nitrogen is detected in the optical, infrared or ultraviolet 
spectra of WC stars. Early recombination line studies of WC stars indicated a
trend of increasing C/He from late-  to early- WC subtypes (Smith \& Hummer 1988\nocite{SH88}) 
with C/He=0.04--0.7 by number, clearly indicative of the exposition of He--burning products. 
 More recent, non-LTE model studies indicate C/He=0.1--0.5
by number, with no such subtype trend (Koesterke \& Hamann 1995). 
Indeed, LMC WC stars possess similar surface abundances to Milky Way counterparts
Crowther et al. (2002a).  Oxygen abundances are more difficult to determine, 
although values rarely exceed O/H=0.1 (Dessart et al. 2000; Crowther et al. 2002a).
Neon is predicted to be enhanced in WC atmospheres, as supported by results based
on mid-IR fine-structure lines from ISO (Dessart et al. 2000), once the clumped 
nature of W-R winds are taken into account.
WO stars are extremely carbon and oxygen rich, as deduced by recombination 
analyses (Kingsburgh et al. 1995) and non-LTE models (Crowther et al. 2000),
with C/He$\geq$0.5 and O/He$\geq$0.1 by number.



\subsection{Slash stars}


All WN9--11 stars discovered to date still possess hydrogen at their surface, 
albeit substantially reduced, with H/He=1.2--3.5 by number for those
optically visible, and H/He$\leq$2 for Galactic Centre He\,{\sc i} sources (Najarro
et al. 1997a).
These also exhibit partially processed CNO material 
from both
stellar (Crowther et al. 1995a; 
Crowther \& Smith 1997) and nebular (Smith et al. 1998) studies.
Abundances patterns of O3If/WN stars mimic those of hydrogen rich WNha stars, 
in the sense that H is little depleted, if at all, whilst CNO elements
are partially processed (Heap et al. 1991; de Koter et al. 1997).

\subsection{Luminous Blue Variables}

Crowther (1997) compares surface He-contents in five LBVs,
revealing a surprisingly uniform mass fraction of H/He=2--2.5
by number. Najarro et al (1997b) has obtained H/He$\sim$3 from
a detail investigation into  P Cygni.
CNO abundances in LBVs have been rarely studied directly 
(e.g Lennon et al. 1997).
More frequently, these are studied via analysis of the
N and O abundances from their associated \HII regions
(Smith et al. 1998). Lamers et al. (2001) discuss evolutionary implications 
for  CNO abundances determined in LBV nebulae.

Most recently, high quality UV and
optical spectra of $\eta$~Car have been obtained with HST and analysed
to reveal enriched nitrogen by at least a factor of ten, whilst
carbon and oxygen are depleted (Hillier et al. 2001).

\section{W-R stars in Local Group galaxies} 

The presence of strong emission lines in W-R stars
 permits their discovery at large distances 
via use of suitable narrow-band filters. 
For example, a set of narrowband filters centred at $\lambda$4650, $\lambda$4686 and a nearby continuum
region will discriminate between WC, WN and normal stars. Groups led by
P. Massey,  and by A.F.J. Moffat in the 1980's used such filter
combinations to find suitable W-R candidates in the Milky Way and
other Local Group galaxies. Suitable filter sets on large 8-10m telescopes
now facilitates identification of W-R stars beyond the Local Group (e.g.
Schild et al. 2003).

\subsection{Visibly obscured W-R stars within the Milky Way}

Since massive stars are solely formed within the disk of the Milky Way,
only a small number can be observed at optical or ultraviolet wavelengths.
At present, little more than 200 W-R stars are known within our Galaxy,
yet many thousand are expected to be present (van der Hucht 2001).
It is only via observation at IR wavelengths where interstellar extinction
is much reduced that one can detect individual stars. 
We have already discussed the young massive
clusters close to the centre of our Galaxy, which contain hot and cool
slash stars, plus several normal W-R stars. Other Galactic
clusters, of which Westerlund 1 is a beautiful example, have recently been
found to host large W-R populations. Clark \& Negueruela (2002) have
identified at least 11 new W-R stars in Westerlund 1, plus a wide
diversity of blue, yellow and red supergiants, such that it has a
mass comparable with any young cluster in the Local Group.

Near-IR surveys, using filters
selected to measure W-R emission lines and continuum regions,
are presently underway (e.g. Homeier et al. 2003).
Real progress awaits application of this method using large format
detectors mounted on large 4m-class telescopes. 

\begin{figure}
\includegraphics[height=7.3cm]{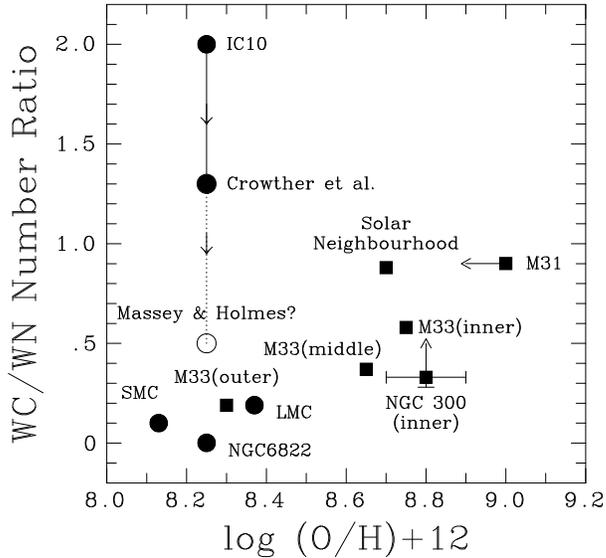}
\caption{WC/WN ratio for Local Group and Sculptor Group 
spiral (squares) and irregular  galaxies (circles) versus oxygen content.
Individual data points are taken from Massey \& Johnson (1998), Schild et al.
(2003), except for IC~10 for which recent spectroscopic (Crowther et al. 2003,
filled circle) and photometric (Massey \& Holmes 2002, open circle) results are 
indicated.}
\end{figure}

\subsection{WR subtype distribution -- the puzzling case of IC~10}

It is well known that the
distribution of W-R stars, specifically the ratio of W-R to O stars,
is a strong function of environment. One can understand this general
behaviour from the metallicity dependence of wind strengths prior
to the W-R phase, in the sense that weaker winds at low metallicies,
such as in the SMC, cause lower mass-loss during the main and immediate
post-main sequence, such that higher initial mass is needed for the
formation of a W-R star via single star evolution (Meynet \& Maeder 1994).

In addition to the reduced W-R/O fraction at lower metallicity, it
is also well known that the number of WC stars to WN stars is also a
sensitive function of metallicity in Local Group galaxies. In the 
Milky Way, WC/WN=3/4 (van der Hucht 2001), 
versus WC/WN=1/10 in the SMC (Massey \& Duffy 2001).
Fig.~2 compares WC/WN ratios with metallicity for various Local Group
and Sculptor Group galaxies.

Late WC subtypes are restricted to metal-rich environments, such
as the inner Milky Way and M31, with
early WC stars found predominantly in metal-poor regions, such as the
Magellanic Clouds. 
Crowther  et al. (2002a) attributed this difference to weaker WC
winds in metal poor regions, since the C\,{\sc iii} $\lambda$5696
classification line is very sensitive to wind density.

Massey (1996) first
highlighted the apparent anomaly of IC~10, which is unique amongst
Local Group galaxies, in the sense that it is metal weak, but has a
large number of WR stars, plus an uniquely
high WC/WN ratio of $\sim$2 (Massey \& Johnson 1998). Recent 
spectroscopy of candidates that had been
identified by Massey et al. (1992)
and Royer et al. (2001) confirmed a high WC/WN ratio of 1.3 (Crowther
et al. 2003). All WC stars were early subtypes, with one exception,
contradicting photometric claims of several late WC stars in IC~10
(Royer et al. 2001).

Recently, a new imaging survey has been carried out
by Massey \& Holmes (2002) suggesting that IC~10 possesses a
normal WC/WN ratio for its metallicity, such that many dozens
of (principally) WN stars had not been detected by earlier surveys.
Of course, spectroscopic follow up is needed to verify the nature
of new candidates. Whatever the case,
 it appears that IC~10 has an exceptional massive star content
and represents the closest galaxy-wide starburst.

W-R stars have been observed in very metal deficient star forming galaxies,
such as I~Zw~18 (de Mello et al. 1998), and also also seen in 
in star forming galaxies at high redshift (Shapley et al. 2003).







%


%

%

\section{Cool Supergiants}

Up to now, this article has dealt with early-type massive stars.
Many, if not all, massive stars briefly pass through a yellow (FG) 
or red (KM) supergiant phase, of which Betelgeuse, 
$\alpha$ Ori (M2.2 Iab) is the best known example. Moderately massive stars
end their life in the red as a Type II supernova (see later), whilst
very massive stars either circumvent the cool supergiant phase entirely, or
subsequently return to the blue as a W-R star. Surprisingly few clusters
are known in which both red supergiants (RSG) and W-R stars are known
to co-exist. Collinder 228 and Berekley 87 each contain one RSG and W-R star
(Massey et al. 2001). Remarkably,
Westerlund~1 hosts at least 2 RSG and 11 W-R stars (Clark \& Negueruela 2002).

Mass-loss in cool supergiants is episodic and involves slow, dense outflows.
Radial pulsations
are generally assumed to provide the mechanism of initiating mass-loss in
M supergiants, as evidenced from circumstellar dust shells (Bowers et al
1983), with some notable exceptions -- VY CMa (M5 Ia) and $\rho$ Cas (F8 Ia+).
Outflow velocities of RSG are typically 10--40 \kms (Jura \& Kleinmann 1990).

Since were are concerned with massive stars here, we shall restrict our
discussion to the most luminous `hypergiants' (de Jager 1998) with 
$\log (L/L_{\odot})> 5$. Studies have largely focussed on M
supergiants, with only highly unusual FG supergiants studied in detail.
IRC+10420 in particular is widely interpreted as a RSG rapidly moving
blueward (Jones et al. 1993; Humphreys et al. 2002), which has recently experienced huge variability
in mass-loss, 10$^{-2}$ to 10$^{-4}$ \Myr, via analysis of its circumstellar
dust shell (Bl\"{o}cker et al. 1999).

Various techniques have been used to quantify mass-loss in late-type supergiants,
whose wind driving mechanism remains uncertain, although these remain much
more controversial than for early-type stars. For example, recent mass-loss
estimates for VY CMa span a factor of twenty -- 1.6$\times 10^{-5}$ \Myr (Josselin
et al. 2000) to 3.5$\times 10^{-4}$ \Myr (Stanek et al. 1995).

Most widely, sub-mm CO emission or mid-IR dust emission permit mass-loss
estimates for cool supergiants.  Jura \& Kleinmann (1990) and Reid et al. 
(1990) have estimated mass-loss rates of Galactic and LMC RSG, respectively,
in which the LMC sample formed a  tight group with $\sim 10^{-4.5}$ \Myr
and the Galactic sample spanned a wide range from 10$^{-6}$ to 10$^{-4}$ \Myr.
More recently, Sylvester et al. (1998) obtained {\it higher} mass-loss rates
of 10$^{-5}$ to 10$^{-4}$ \Myr for Galactic M supergiants using
the 10$\mu$m silicate emission feature, whilst van Loon et al. (1999) have obtained 
LMC RSG mass-loss rates of $\leq 10^{-5}$ \Myr, {\it lower} than Reid et al.
Therefore, depending on the studies used, LMC supergiants
either exceed those in the Milky Way, or vice versa. It is clear that much
research remains to be undertaken in order to firmly establish mass-loss in 
late-type supergiants.





\section{End states of massive stars}

\subsection{Supernovae}

All supernovae (SN), with the exception of Type Ia's, originate from
the core collapse of
massive stars.  Those ending their life as red supergiants explode as
a Type II SN, although in some rare cases, Type II SN progenitors are
blue supergiants, such as Sk$-$69$^{\circ}$ 202, the B3\,I 
progenitor of SN1987A. The origin of this remains unclear,
although the most credible scenarios involve binarity.
Stars which were initially even more massive 
end their life as either a Type Ib or Ic SN. These 
are  presumably related to  WN and WC progenitors, respectively,
given that neither possess spectral lines of hydrogen, whilst
 helium is also absent in Type Ic's. 
Considering all galaxy types in the Local
Universe, Cappellaro et al. (1999) found that 70\% of all observed
SN are due to core collapse massive stars, predominantly Type II.


Light curve and spectroscopic analysis of SN permit ejecta
masses to be determined. Iwamoto et al. (1998, 2000) derived ejecta
CO masses of 2--14$M_{\odot}$ for the Type Ic supernovae SN~1994I,
SN~1997ef and SN~1998bw. The masses discussed here represent minimum masses for the final
progenitor, since an unknown (model dependent) fraction would have
imploded to form a black hole or other compact object. Nevertheless,
the ejecta masses of SN~1997ef and SN~1998bw fit in well with the final
masses of LMC WC stars derived by Crowther et al. (2002a).

SN~1998bw possessed exceptional
wind velocities (27\,000 \kms) immediately after the explosion, such 
the energy of the ejecta exceeded 10$^{52}$ erg. Consequently, it
has been denoted a `hypernova', an exceptional class of energetic
SN.   Of course,
the large explosion energy claimed for SN~1998bw is based on the assumption
was spherically symmetric. There is growing evidence that
SN Ib/c show a small polarization, implying that their explosions might
be somewhat aspherical. 

Other evidence linking Type Ib/c SN to Wolf-Rayet precursors includes
pre-SN imaging of M74, the host galaxy of SN~2002ap (Type Ic SN), which led Smartt
et al. (2002) to conclude that a Wolf-Rayet star was a 
viable candidate for the immediate progenitor..


\subsection{Hypernovae and Gamma Ray Bursts}

It is only within the past few years that Gamma Ray Bursts (GRBs), 
in which 0.1--1 MeV photons are detected for a  few seconds, have
been shown to be of cosmological origin, typically in the range 
$z$=0.5 to 2. There are two, potentially
distinct populations of GRBs, peaking at 0.2s and 20s (Fishman \& Meegan 1995),
which may have different origins, the latter being most common.

 The rapid (millisecond)  $\gamma$-ray variability of GRBs 
implies a compact object, of size smaller than a light millisecond
(10$^7$ cm) and mass not greater than $\sim$30 $M_{\odot}$. The total energy
released in $\gamma$-rays corresponds to a rest mass energy of a fraction of a solar mass.
The most natural way for triggering a GRB event is the accretion of a fraction of a solar mass onto a 
black hole. Therefore, most cosmological GRB models have at their basis gravitational
collapse of a (several) solar mass progenitor to a black hole. The  duration of long bursts
implies that some fraction of the progenitor mass accretes via a disk. The so-called 
`collapsar' model for long duration bursts, favours a rapidly rotating (massive) stellar core 
which collapses to form a  
rotating black hole plus an accretion torus (Woosley et al. 1999).

The first direct link between core collapse SNB and GRBs was established by the
`hypernova' SN~1998bw which was spatially coincident with 
GRB\,980425 (Galama et al. 1998). The host galaxy
for this source was established to lie at $z$=0.0085, 
such that  the GRB was found to be intrinsically
many orders of magnitude fainter than classical GRBs
(Woosley et al. 1999).  Consequently, the link 
between `hypernovae' and GRBs still awaited more general confirmation.
Indeed, there are some hypernovae for which no GRB counterpart has been proposed,
including SN~2002ap which was less energetic than SN~1998bw. It is possible that 
weaker explosions are less efficient in collimating the $\gamma$-rays to give rise to
a GRB, or there is an inclination of the beam axis with respect to the 
line-of-sight to the SN.

Subsequently,  Reeves et al. (2002) observed the 
X-ray afterglow of GRB\,011211 using XMM-Newton, revealing
spectral lines exhibiting an outflow velocity of 0.08c, plus evidence of
 enriched $\alpha$-group elements, characteristics 
unique to core collapse SN. However the GRB was too faint for the SN to be observed
optically. 

Very recently, the extremely bright, long duration GRB\,030329  
(Stanek et al. 2003), at a low redshift of $z$=0.1685, has been found to
reveal a `hypernova' spectrum SN~2003dh. Other cases in which supernovae
has been inferred from light curves and colours of GRB afterglows,
though this represented direct proof that a subset of classical GRBs 
originate from SN. From the accumulated
evidence at hand, it is clear that there is a direct
link between SN originating from core collapse massive stars and
long duration GRBs. 

\section{Further Reading}

Current research relating to Wolf-Rayet stars can be found in
the proceedings of IAU Symposium 212 (van der Hucht et al. 2003),
whilst Luminous Blue Variables are discussed in Nota \& Lamers (1997).
A recent volume in the Lecture Notes in Physics series discusses
Supernova and GRBs (Weiler 2003).




\end{document}